%% file: dualgen.tex
\documentclass[a4paper,12pt]{article}
\usepackage{epsfig}
\usepackage{psfig}

\begin{document}

\begin{titlepage}

\baselineskip 24pt

\begin{center}

{\Large {\bf Yang-Mills Duality and the Generation Puzzle}}\\

\vspace{.5cm}

\baselineskip 14pt

{\large CHAN Hong-Mo}\\
chanhm\,@\,v2.rl.ac.uk  \,\,\,  \\
{\it Rutherford Appleton Laboratory,\\
  Chilton, Didcot, Oxon, OX11 0QX, United Kingdom}\\

\end{center}

\vspace{.3cm}

\begin{abstract}

The fermion generation puzzle has survived into this century as one of 
the great mysteries in particle physics.  We consider here a possible 
solution within the Standard Model framework based on a nonabelian 
generalization of electric-magnetic duality.  First, by constructing
in loop space a nonabelian generalization of the abelian dual transform 
(Hodge *), one finds that a ``magnetic'' symmetry exists also in classical 
Yang-Mills theory dual to the original (``electric'') gauge symmetry.  
Secondly, from a result of 't~Hooft's, one obtains that for confined 
colour $SU(3)$, the dual symmetry $\widetilde{SU}(3)$ is spontaneously 
broken and can play the role of the ``horizontal symmetry'' for generations.  
Thirdly, such an identification not only offers an explanation why there 
should be three and apparently only three generations of fermions with 
the remarkable mass and mixing patterns seen in experiment, but allows 
even a calculation of the relevant parameters giving very sensible results.  
Other testible predictions follow ranging from rare hadron decays to 
cosmic ray air showers.

\end{abstract}

\end{titlepage}

\clearpage

Over the last century, giant steps were made in our understanding of the
fundamental structure of the physical world culminating in the so-called
Standard Model which seems to cover at present every known experimental
fact.  And the whole is based gratifyingly on a very beautiful framework,
namely that of the Yang-Mills Theory, which is itself a generalization 
of the gauge principle discovered earlier in Maxwell's theory of 
electormagnetism.    

One very puzzling question which remains, however, is why there should 
be three and apparently only three generations of fermions, a fact which
is simply taken for granted in the Standard Model.  As far as we know 
today, our world is built out of fundamental fermions of the following 
twelve types:
\begin{equation}
\left( \begin{array}{c} t \\ c \\ u \end{array} \right); \ \ 
   \left( \begin{array}{c} b \\ s \\ d \end{array} \right); \ \ 
   \left( \begin{array}{c} \tau \\ \mu \\ e \end{array} \right); \ \ 
   \left( \begin{array}{c} \nu_1 \\ \nu_2 \\ \nu_3 \end{array} \right).
\label{fermgens}
\end{equation}
The quarks in the first two columns are distinguished from the leptons in the
last two by the quarks having colour but leptons not, while the up- and 
down-quarks, as with the charged leptons and neutrinos, are distinguished
by their different weak isospins.  Thus, in a sense, one can understand why
Nature would want this variety for building her multifarious universe.  But 
why should she want three copies for each colour and weak isospin?  As far as 
we know, these three copies, called generations, are distinguished only by 
their masses and these themselves fall into a very remarkable pattern.  For
the first three charged fermion-types, they are in MeV units roughly 
\cite{databook}:
\begin{equation}
m_t = 180000, \ \ m_c = 1200, \ \ m_u = 4;
\label{Umasses}
\end{equation}
\begin{equation}
m_b = 4200, \ \ m_s = 120, \ \ m_d = 7;
\label{Dmasses}
\end{equation}
\begin{equation}
m_\tau = 1800, \ \ m_\mu = 100, \ \ m_e = .5,
\label{Lmasses}
\end{equation}
dropping from generation to generation by one to two orders of magnitude, 
a phenomenon referred to in the trade as the ``fermion mass hierarchy''.
(Presumably, the masses of the three neutrinos would follow a similar 
pattern but of this we are not yet certain because of the experimental 
difficulty in measuring the very small masses of these neutral particles.)

Nor does the mystery stop there.  The state vectors representing the
three generations are approximately but NOT exactly aligned between the
different fermion-types.  Suppose we were to represent the three states
of each fermion-type by a orthonormal triad in generation space, and the
relative orientation of the down-triad to the up-triad by a unitary
matrix, known in the trade as the CKM matrix for quarks and the MNS matrix
for leptons, present experiment give approximately for the absolute values
of the matrix elements \cite{databook}:
\begin{equation}
|V_{CKM}| = \left( \begin{array}{lll} 0.975 & 0.220 & 0.003 \\
                                    0.220 & 0.974 & 0.04  \\
                                    0.008 & 0.04  & 0.999 \end{array} \right),
\label{CKMmat}
\end{equation}
\begin{equation}
|U_{MNS}| = \left( \begin{array}{ccc} ? & 0.4 - 0.7 & 0.0 - 0.15 \\
                                    ? & ? & 0.45 - 0.85 \\
                                    ? & ? & ? \end{array} \right),
\label{MNSmat}
\end{equation}  
where we have ignored in each matrix a $CP$-violating phase for which little
yet is known.  One sees that the matrix for quarks is tantalisingly close 
to but definitely not the identity, with the nonzero off-diagonal elements 
representing the rates of some very well measured hadronic proscesses.  
Whereas for the leptons, the matrix is far from diagonal with the large 
off-diagonal elements\footnote{The quoted bounds for $U_{12}$ correspond
to either the large angle MSW or the vacuum oscillation solution, but not
the small angle MSW solution, in solar neutrinos.} representing the recent 
results from some beautiful well-publicized experiments on neutrino 
oscillations \cite{SuperK,Chooz}.

Why should there be three fermion generations and why should they fall into
such intriguing mass and mixing patterns?  This question is what is meant 
by the ``generation puzzle'' which has been plaguing particle theorists in 
different forms for over half a century, say ever since Feynman reputedly 
pasted over his bed the question ``Why does the muon weigh?''  Even if one 
is not worried by deeper questions of whys and wherefores, the question 
still represents in practical terms a large number of empirical parameters.  
In what we now call our Standard Model of particle physics, all the quantities
listed in (\ref{Umasses}) - (\ref{MNSmat}) have to be fed in from experiment 
and account together for nearly three-quarters of the total number of 
parameters required to define the Model.  Hence, the lack of an explanation 
for the generation puzzle not only reduces considerably the Model's predictive
power but also subtracts from our confidence in its fundamentality.  It is 
thus no wonder that the puzzle is regarded by many as one of the most urgent 
now facing particle physicists.  
   
A popular and seemingly reasonable approach to the generation puzzle is to
postulate a new (``horizontal'') 3-fold broken symmetry to account for 
the three generations, which still begs of course the questions first, why 
it is 3 and not some other number, and second, where this horizontal symmetry 
comes from.  One can try to look for answers in larger theories which 
contain the Standard Model, such as grand unified theories or strings, but 
this usually involves introducing more freedom and reduces the predictive 
power on the generation puzzle itself.  What I wish to do here instead is 
to explore with you a, to me, attractive alternative, namely an explanation 
for the puzzle from within the Standard Model framework.  I hope to show 
you that this proposition is not as difficult as it might sound at first 
sight, for the gauge principle as embodied in the Yang-Mills Theory is such 
a beautifully rich construct that it actually admits within the Standard 
Model such an horizontal symmetry.  And this symmetry is able not only 
to explain why there should be three and only three generations with mass 
and mixing patterns similar to those noted above, but even to allow the 
calculation of the relevant parameters giving quite sensible answers.

To explain the idea simply, let me go all the way back to classical 
electromagnetism.  Here at any point in space-time free of electric and 
magnetic charges, the field tensor $F_{\mu\nu}$ satisfies the equations:
\begin{equation}
\partial^\nu F_{\mu\nu}(x) = 0,
\label{Maxwelleq}
\end{equation}
and
\begin{equation}
\partial^\nu *F_{\mu\nu}(x) = 0,
\label{Gausslaw}
\end{equation}
where $*F_{\mu\nu}$ is the dual field:
\begin{equation}
*F_{\mu\nu} = - \frac{1}{2} \epsilon_{\mu\nu\rho\sigma} F^{\rho\sigma}
\label{Hodgestar}
\end{equation}
As a consequence of these equations, there exist locally a potential to 
both $F_{\mu\nu}$ and $*F_{\mu\nu}$, thus:
\begin{equation}
F_{\mu\nu}(x) = \partial_\nu A_\mu(x) - \partial_\mu A_\nu(x),
\label{defAmu}
\end{equation}
\begin{equation}
*F_{\mu\nu}(x) = \partial_\nu \tilde{A}_\mu(x) - \partial_\mu \tilde{A}_\nu(x).
\label{defAmut}
\end{equation}
In other words, both $F_{\mu\nu}$ and $*F_{\mu\nu}$ are gauge fields.
In terms of the gauge field $F_{\mu\nu}$, electric charges appear as
sources while magnetic charges appear as monopoles.  But in terms of the
gauge field $*F_{\mu\nu}$, magnetic charges appear as sources while electric
appear as monopoles.  Next, under the transformations:
\begin{equation}
A_\mu(x) \longrightarrow A_\mu(x) + \partial_\mu \alpha(x),
\label{gaugetrans}
\end{equation}
\begin{equation}
\tilde{A}_\mu(x) \longrightarrow \tilde{A}_\mu(x) + \partial_\mu 
   \tilde{\alpha}(x),
\label{gaugetranst}
\end{equation}
where $\alpha$ and $\tilde{\alpha}$ are independent functions of $x$, both 
$F_{\mu\nu}$ and $*F_{\mu\nu}$ are invariant, which means that the theory has
a doubled gauge symmetry, say $U(1) \times \tilde{U}(1)$.  Notice that this
does not mean doubled physical degrees of freedom since $F_{\mu\nu}$ and
$*F_{\mu\nu}$ are still related by (\ref{Hodgestar}) and represent still
the same physical degrees of freedom.  That classical electromagnetism 
has these ``dual'' properties is of course well known.  The only reason 
why this doubled symmetry has so far not played a significant role is just 
the experimental fact that no magnetic charge has yet been found.

Given electric-magnetic duality in the abelian theory, it is natural to 
ask whether the concept is generalizable to nonabelian Yang-Mills fields.
If we were to keep the dual transform as the Hodge star (\ref{Hodgestar}),
then the answer is no.  The problem is that, in contrast to the abelian 
theory, the dual $*F_{\mu\nu}$ to the Yang-Mills field $F_{\mu\nu}$ is 
not in general a gauge field derivable from a gauge potential.  The field 
$F_{\mu\nu}$, of course, is by assumption a gauge field derivable from a
potential thus:
\begin{equation}
F_{\mu\nu}(x) = \partial_\nu A_\mu(x) - \partial_\mu A_\nu{x} 
   + ig [A_\mu(x), A_\nu(x)],
\label{FmunuinA}
\end{equation}
and satisfies, at points of space-time free of ``colour'' charges, the 
equation:
\begin{equation}
D^\nu F_{\mu\nu}(x) = 0,
\label{Yangmillseq}
\end{equation}
in close analogy to (\ref{Maxwelleq}) apart from the replacement of the 
partial by the covariant derivative $D_\nu$.  However, in contrast to 
the abelian case where (\ref{Maxwelleq})ensures the local existence of the 
dual potential $\tilde{A}_\mu$, the equation here (\ref{Yangmillseq}) does 
not offer the same guarantee.  In fact, counter-examples \cite{Guyang} are
known of fields $F_{\mu\nu}$ satisfying (\ref{Yangmillseq}) for which no 
potential $\tilde{A}_\mu$ exists for $*F_{\mu\nu}$. 

\begin{figure}
\centering
\includegraphics{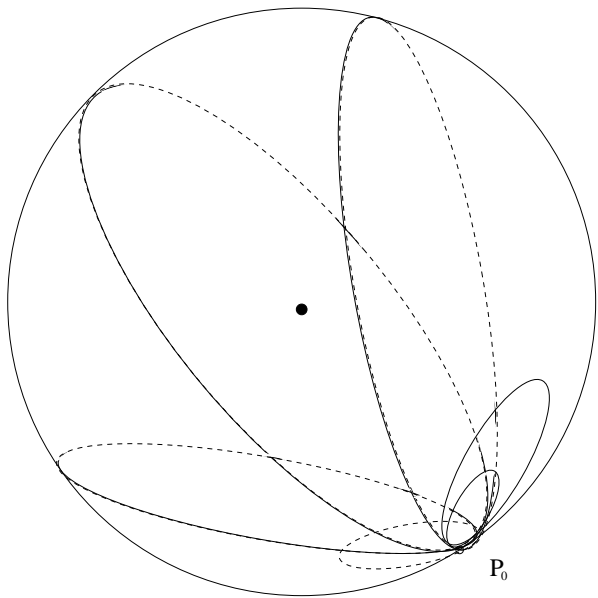}
\caption{Definition of a nonabelian monopole}
\label{looptheloop}
\end{figure}

However, this conclusion does not by itself rule out the possiblity that
by defining the dual transform differently for nonabelian Yang-Mills fields, 
though reducing back to (\ref{Hodgestar}) in the abelian case, one may be able
to recover dual properties as a result.  Indeed, we think we have succeeded 
in doing so by proceeding as follows \cite{dualsymm}.  First, we asked 
ourselves the question whether there is a condition on the nonabelian field
similar to the abelian condition (\ref{Gausslaw}) which guarantees the 
existence of a potential.  This is not obvious.  Although a similar 
equation:
\begin{equation} 
D_\nu *F_{\mu\nu}(x) = 0
\label{Bianchi}
\end{equation}
holds as Bianchi identity for any field satisfying (\ref{FmunuinA}), the 
converse is not true, meaning that the condition (\ref{Bianchi}) does not
guarantee the existence of a potential $A_\mu$ for $F_{\mu\nu}$.  To find
instead a condition that does, we reason as follows.  The physical content of
(\ref{Gausslaw}) for the abelian theory is that there is no magnetic charge
or monople for $F_{\mu\nu}$ at the point $x$.  Could it not be that the 
condition required for the nonabelian field should also be the statement 
that it should have no monopole at that point?  To answer this, we recall
first the definition of a nonabelian monopole \cite{Lubkin,Wuyang,Coleman}.
For each loop $C$ in space-time, construct the phase factor:
\begin{equation}
\Phi(C) = P \exp ig \oint_C A_\mu(x) dx^\mu,
\label{PhiC}
\end{equation}
which maps loops in space-time to points in the structure group $G$.  Then
consider the one-parameter family of loops $C_t, t = 0 \rightarrow 2 \pi$ 
enveloping a surface $\Sigma$ enclosing the point $x$, as illustrated in 
Figure \ref{looptheloop}.
As $C_t$ loops over the surface, $\Phi(C_t)$ traces out a closed curve 
$\Gamma_\Sigma$ in the group $G$.  The monopole charge at $x$ (independent 
of surface $\Sigma$ by continuity) is then defined as the homotopy class 
of this curve $\Gamma$ in $G$.  Obviously, if the group $G$ is simply
connected, the theory has no monopole charges, but theories with multiply
connected gauge groups will have monopole charges.  For instance, a theory 
with the doubly-connected $SO(3)$ as gauge group admits monopole charges 
taking the values $\pm$ in $Z_2$, which will serve us later as a simple 
example for illustration.   

The above definition of the monopole charge itself being given in terms of 
loop-dependent quantities, the convenience of a loop space formalism for
our problem is indicated.  To encapsulate therefore this somewhat unwieldy 
definition into a formula, let us adopt a formalism suggested by Polyakov 
\cite{Polyakov} and introduce the connection in loop space as field variable.
We choose now also to work in paramatrized loop space, being much the more 
convenient, where the geometric loops $C$ above are parametrized as:
\begin{equation}
C: \{\xi^\mu(s); \ \ s = 0 \rightarrow 2 \pi, \xi^\mu(0) = \xi^\mu(2 \pi)
   = \xi^\mu_0 \},
\label{ximus}
\end{equation}
so that loop-dependent quantities such as $\Phi(C)$ are now just functionals
$\Phi[\xi]$ of the function $\xi(s)$:
\begin{equation}
\Phi[\xi] = P_s \exp ig \int_0^{2 \pi} ds A_\mu(\xi(s)) \frac{d\xi}{ds}.
\label{Phixi}
\end{equation}
and loop derivatives are just functional derivatives.  The loop connection 
which is chosen as field variable we denote as $F_\mu[\xi|s]$, following 
Polyakov.  In case a potential $A_\mu(x)$ exists, then $F_\mu[\xi|s]$ is 
expressible as the logarithmic derivative of of the phase factor, namely:
\begin{equation}
F_\mu[\xi|s] = \frac{i}{g} \Phi^{-1}[\xi]\frac{\delta}{\delta \xi(s)} 
   \Phi[\xi].
\label{Fmuxis}
\end{equation}
What we seek is the converse, namely a condition on $F_\mu[\xi|s]$ to recover
a potential $A_\mu(x)$ in terms of which $F_\mu[\xi|s]$ is expressible via
(\ref{Fmuxis}) and (\ref{Phixi}).

\begin{figure}
\centering
\input{loopcurv.pstex_t}
\caption{illustration for the loop space curvature $G_{\mu\nu}[\xi|s]$}
\label{skiprope}
\end{figure}
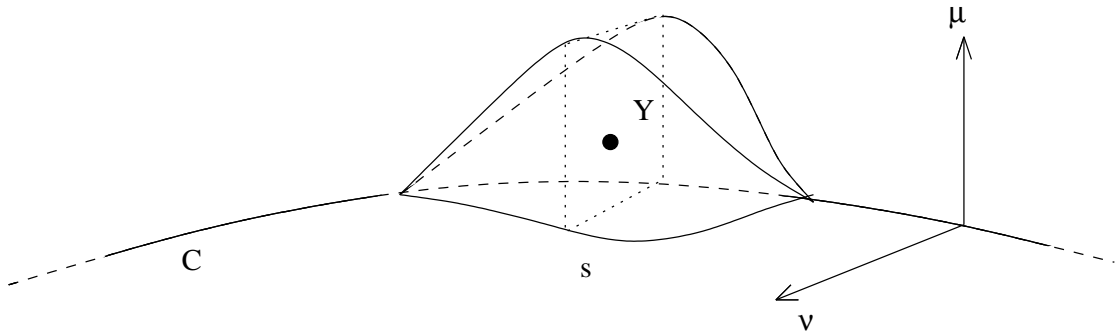

With the connection in loop space as variable, a monopole charge as defined
above can be simply expressed as a nontrivial holonomy, or equivalently, in 
differential terms, as a nonvanishing loop space curvature \cite{Chanstsou1,
Chanstsou2,Chantsoubk}.  Explicitly. in terms of $F_\mu[\xi|s]$, the condition 
that there is an $SO(3)$-monopole with charge $-$ on the world-line $Y(\tau)$ 
can be written as:
\begin{equation}
G_{\mu\nu}[\xi|s] = -4 \pi \tilde{g} \int d\tau \kappa[\xi|s] 
   \epsilon_{\mu\nu\rho\sigma} \frac{d Y^\rho(\tau)}{d\tau}
   \frac{d \xi^\sigma(s)}{ds} \delta(\xi(s) - Y(\tau)),
\label{Gausslawq}
\end{equation}
with:
\begin{equation}
\exp (i \pi \kappa) = -1,
\label{kappa}
\end{equation}
where $G_{\mu\nu}$ is the loop space curvature defined as:
\begin{equation}
G_{\mu\nu}[\xi|s] = \frac{\delta}{\delta \xi^\nu(s)} F_\mu[\xi|s]
                    - \frac{\delta}{\delta \xi^\mu(s)} F_\nu[\xi|s]
                    + ig [F_\mu[\xi|s], F_\nu[\xi|s]].
\label{Gmunuxis}
\end{equation}
In ordinary space-time, $G_{\mu\nu}[\xi|s]$ can be visualized as in 
Figure \ref{skiprope}
where the loop skips over a little 3-volume enclosing the point $\xi(s)$, so 
that (\ref{Gausslawq}) is clearly seen to be essentially just a differential 
version of the above definition of the monopole charge as a homotopy class 
in terms of a finite surface $\Sigma$.

Having obtained an expression for the monopole charge, we can now write 
down the monopole-free condition as:
\begin{equation}
G_{\mu\nu}[\xi|s] = 0,
\label{nomonopole}
\end{equation} 
which we thought might be the guarantee we sought for the existence of the
gauge potential $A_\mu(x)$.  And indeed, apart from some minor technical 
conditions, this was shown to be just what was needed to recover $A_\mu(x)$ 
from $F_\mu[\xi|s]$ \cite{Chanstsou1}.  Notice that, in parallel to the
condition (\ref{Gausslaw}) for the abelian theory, (\ref{nomonopole}) 
implies the local existence of $A_\mu(x)$ independently of whether monopole
charges occur at other points of space-time and can thus be used to recover
the potential everywhere except at the (isolated) locations of monopoles.

With this result in hand, we return to the question of duality for nonabelian 
theories.  For this to work, we need a potential for the dual field as well.
If we represent the dual field by $\tilde{F}_\mu[\xi|s]$ the relation of
which to the original field $F_\mu[\xi|s]$ is yet to be discovered, then we 
would like $\tilde{F}_\mu[\xi|s]$ as a loop space connection to satisfy 
the zero-curvature condition, namely:
\begin{equation}
\tilde{G}_{\mu\nu}[\xi|s] = 0,
\label{nosource}
\end{equation}
where the dual loop space curvature $\tilde{G}_{\mu\nu}[\xi|s]$ is defined 
as in (\ref{Gmunuxis}) but with $\tilde{F}_\mu[\xi|s]$ as the connection.  
Recall now that in the abelian theory, the reason duality worked was that 
the source-free condition (\ref{Maxwelleq}) for the field $F_{\mu\nu}(x)$ 
saying that there is no electric charge at the point $x$ in space-time 
coincides with the condition required to ensure the existence of the dual 
potential $\tilde{A}_\mu(x)$ at the same point.  Hence, to obtain a similar
result for the nonabelian theory, we can hope to construct a dual transform 
such that the source-free condition on $F_\mu[\xi|s]$ saying that there is 
no ``colour'' charge at $x$, which according to Polyakov \cite{Polyakov} 
reads in loop space as:
\begin{equation}
\frac{\delta}{\delta \xi^\mu(s)} F^\mu[\xi|s] = 0,
\label{Polyakoveq}
\end{equation}
should coincide with the condition (\ref{nosource}) for ensuring the
existence of the dual potential $\tilde{A}_\mu(x)$.  In this we think we 
have succeeded \cite{dualsymm} with the dual transform:
\begin{equation}
\omega^{-1}(\eta(t)) \tilde{E}_\mu[\eta|t] \omega(\eta(t))
   = -\frac{2}{\bar{N}} \epsilon_{\mu\nu\rho\sigma} \dot{\eta}^\nu(t)
     \int \delta \xi ds E^\rho[\xi|s] \dot{\xi}^\sigma(s) \dot{\xi}^{-2}(s)
     \delta(\xi(s) - \eta(t)),
\label{Dualtrans}
\end{equation}
given in terms of the variables $E_\mu[\xi|s]$ and $\tilde{E}_\mu[\xi|s]$
which are closely related to the previous variables $F_\mu[\xi|s]$ and its
dual $\tilde{F}_\mu[\xi|s]$.  It will take too long to explain in detail 
the meaning of the various symbols entering into the above formula and I
shall not do so, but refer the interested reader to our papers.  I should 
stress that, involving as they do rather delicate operations in loop space, 
neither the proposed dual transform (\ref{Dualtrans}) nor the conclusion 
deduced from it can lay any claim to mathematical rigour.  Barring this 
reservation, however, we think we have found a generalization of classical 
electric-magnetic duality to nonabelian Yang-Mills fields with the desired
properties.  In particular, this implies that, in analogy to the abelian 
theory, Yang-Mills theory has also a doubled gauge symmetry, say e.g.
$SU(N) \times \widetilde{SU}(N)$, where the two $SU(N)$'s are identical as
groups but differ by parity, with the first $SU(N)$ representing (electric)
colour and the second $\widetilde{SU}(N)$ dual (magnetic) colour.  As in
the abelian theory also, this doubling of the symmetry does not mean a
doubling of the physical degrees of freedom; the theory can be described
by either of the two dual sets of field variables, say by either $A_\mu$ or
$\tilde{A}_\mu$, not both.  Furthermore, it can be shown that in terms of 
$A_\mu$, colour charges appear as sources and dual colour charges as 
monopoles, while in terms of $\widetilde{A}_\mu$, the reverse is true 
\cite{predualsymm}.

As should be obvious from the discussion, the generalized duality outlined
above is attained in full only in the classical Yang-Mills theory, which by 
itself has little scope for physical application.  And the formulation being 
in loop space and already very complicated, we have at present little idea 
how the theory can be quantized.  Fortunately, however, one property of the 
quantum theory is known, which when coupled with an old result of `t~Hooft
\cite{tHooft} leads to a conclusion highly suggestive for the existence of
generations.  This comes about as follows.  Taking the (group) trace of the
phase factor (\ref{PhiC}), one has for the quantized theory the Wilson
operator:
\begin{equation}
A(C) = {\rm Tr} P \exp ig \oint_C A_\mu(x) dx^\mu,
\label{AC}
\end{equation}
which in the words of 't~Hooft ``measures (colour) magnetic flux {\it through}
$C$ and creates an (colour) electric flux line {\it along} $C$''.  Given now
from above the dual potential $\tilde{A}_\mu$, one can construct analogously
also the dual operator:
\begin{equation}
B(C) = {\rm Tr} P \exp \ \tilde{g} \oint_C \tilde{A}_\mu(x) dx^\mu.
\label{BC}
\end{equation} 
And if the duality discussed here is the same as the duality studied by
't~Hooft, then $B(C)$ should measure (colour) electric flux through $C$ 
while creating a (colour) magnetic flux line along $C$, hence satisfying
't~Hooft's commutation relation:
\begin{equation}
A(C) B(C') = B(C') A(C) \exp (2 \pi i n/N)
\label{ABcomrel}
\end{equation}
for space-like loops $C$ and $C'$ with linking number $n$ between them.
That this commutation relation indeed holds for $A(C)$ and $B(C)$ as given 
in (\ref{AC}) and (\ref{BC}) was shown in \cite{dualcomm} using the apparatus
developed above and the appropriate Dirac quantization relation between $g$
and $\tilde{g}$.  This means therefore that 't~Hooft's result on confinement 
in \cite{tHooft} applies in the present framework.

In particular, for colour dynamics with the doubled symmetry $SU(3) \times
\widetilde{SU}(3)$, where colour $SU(3)$ is supposed to be confined, 't~Hooft's
result would imply that the dual colour symmetry $\widetilde{SU}(3)$ should
be broken.  In other words, already within the framework of the Standard
Model, the considerations above would automatically lead to the occurence 
of a broken 3-fold symmetry which can play the role of the ``horizontal
symmetry'' demanded by the empirical phenomenon of generations.  Notice that, 
according to the above logic, such a broken $\widetilde{SU}(3)$ would occur
in any case, and if so would lead in principle to observable consequences
which would have to be accounted for eventually.  That being the case, it
seems to us natural to attempt identifying dual colour with generation
and explore the consequences of this bold assumption \cite{dualcons}. 

Apart from offering an immediate explanation for the existence of three,
and apparently only three, generations, the identification of generation
to dual colour in the present scheme has another attractive feature of
even suggesting Higgs fields for breaking the dual colour symmetry.  One 
notices that in the dual transform (\ref{Dualtrans}), there appears a 
rotation matrix denoted by $\omega$ which transforms between frames in
$SU(N)$ and $\widetilde{SU}(N)$.  In the presence of charges, whether
colour or dual colour, this matrix, or equivalently the frame vectors in
$SU(N)$ or $\widetilde{SU}(N)$, will have to be patched \cite{predualsymm},
so that, if we follow the arguments of \cite{Wuyang2}, they will
acquire dynamical roles.  Considered as fields (rather like vierbeins
in gravity) these frame vectors, being space-time scalars belonging to
the fundamental representation of the structure groups, can play very well
the role of Higgs fields.\footnote{The assumption of Higgs fields in the 
fundamental representation means that they are taken to be fundamental 
fields like the Higgs fields in the electroweak theory and not dynamically 
generated from colour or dual colour dynamics.}

Suppose then we make this second bold assumption of identifying frame vectors
with Higgs fields, one obtains for breaking $\widetilde{SU}(3)$ three Higgs
triplets, which being frame vectors with equal status, should (we argue)
appear symmetrically in the action.  This then suggests a Yukawa coupling
of the following form:
\begin{equation}
\sum_{(a)[b]} Y_{[b]} \bar{\psi}_L^a \phi_a^{(a)} \psi_R^{[b]},
\label{Yukawa}
\end{equation}
where, as in electroweak theory, we have taken left-handed fermions in
the fundamental representation, i.e. triplets, and right-handed fermions
as singlets.\footnote{Note that in order to accommodate $\widetilde{SU}(3)$
dual colour triplets as well as $SU(3)$ colour triplets like quarks, it
is essential that colour be embedded in a larger gauge group like that
of the Standard Model.  For an explanation of this point, see for example
\cite{dualcons} and references therein.}  In turn the Yukawa coupling 
(\ref{Yukawa}) implies at tree-level a (hermitized) mass matrix of the 
following factorized form:
\begin{equation}
m = m_T \left( \begin{array}{c} x \\ y \\ z \end{array} \right) (x, y, z),
\label{massmat}
\end{equation}
where $(x, y, z)$ is a normalized vector with its components given by 
the vacuum expectation values of the Higgs fields $\phi^{(a)}$.  It 
follows therefore that at tree-level, $m$ has only one non-zero 
eigenvalue, i.e. one heavy state with the other two massless which 
we may interpret as embryo mass hierarchy, and, $(x, y, z)$ being
the same for all fermion-types, zero mixing between up- and down-states.
This is not a bad zeroth order approximation at least for charged leptons
and quarks.

\begin{figure}
\centerline{\psfig{figure=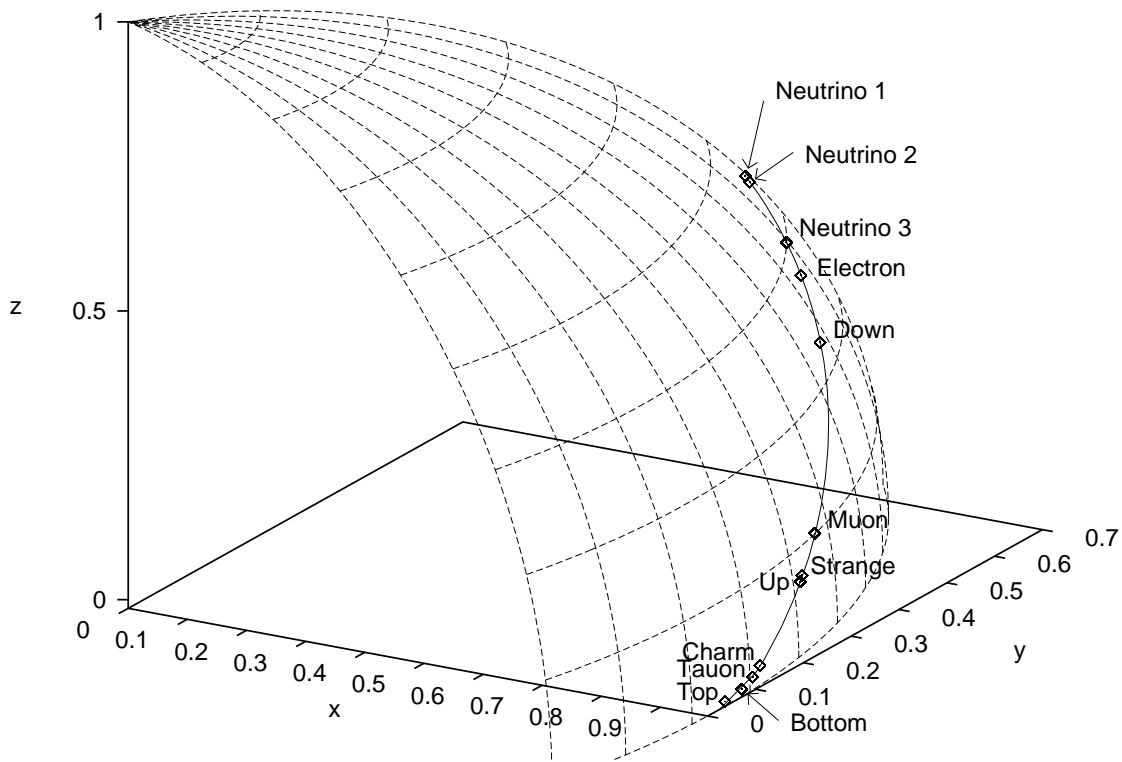,width=0.85\textwidth}}
\caption{Trajectory traced out by the vector $(x, y, z)$ in generation space}
\label{Jakovsphere}
\end{figure}

Under radiative loop corrections, however, the vector rotates with the
energy scale where the rotation depends on the fermion-type, so that up- 
and down states become disoriented with respect to each other leading 
to nontrivial mixing matrices.  At the same time, mass starts to 
``leak'' from the top generation into the two lower generations giving
them small but nonzero masses.  Indeed, a calculation to one-loop level 
\cite{ckm,nuosc,phenodsm}, the details of which need not here 
bother us, yields the following picture.  As the energy changes,
the vector $(x, y, z)$ rotates and traces out a trajectory on the unit 
sphere.  At high energy it starts from near the fixed point $(1,0,0)$ and 
moves, as energy lowers, towards the fixed point $\frac{1}{\sqrt{3}}(1,1,1)$.
Although the trajectories can in principle be different for different
fermion-types, the data demand, for reasons yet theoretically unclear, that 
they coincide to a very good approximation.  The 12 different fermion states
listed above in (\ref{fermgens}) thus only occupy different points on 
this single trajectory.  The actual picture obtained is shown in Figure
\ref{Jakovsphere}.\footnote{Note that this picture obtains for neutrinos
only for the so-called vacuum oscillation solution of the solar neutrino
puzzle.  The ``leakage'' mechanism here can give only a hierarchical 
fermion mass spectrum, and the (Dirac) mass ratio $m_{\nu_2}/m_{\nu_1}$
implied by the MSW solutions for solar neutrinos is just too large to be
accommodated by the scheme, at least in its present form.}

\begin{figure}
\centerline{\psfig{figure=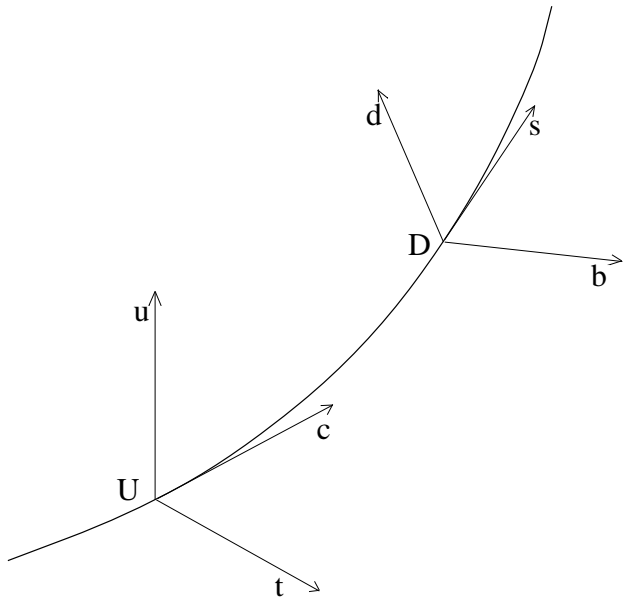,width=0.6\textwidth}}
\caption{The (Darboux) triad of state vectors for the 3 generations}
\label{Darboux}
\end{figure}

It is intriguing that most of the peculiar qualitative features noted before 
in the fermion mass and mixing patterns in (\ref{Umasses})-(\ref{MNSmat}) 
can now be read off immediately from this single picture.  We note first 
that since both fermion mixing and lower generation masses occur as the 
result of the rotation of the vector $(x, y, z)$ with changing scales, 
the slower the rotation, the smaller will be the mixing and the ``leakage'' 
of masses to the lower generations.  Now, the top quark being heavier than 
the bottom and therefore closer to the fixed point $(1,0,0)$ on the 
trajectory, as depicted in Figure \ref{Jakovsphere}, is at a location where 
the rotation is slower.  Hence, we expect that the leakage from the top to 
be less than that from the bottom, giving a much smaller ratio to $m_c/m_t$ 
than to $m_s/m_b$.  Similarly, $m_b$ being large than $m_\tau$, we expect
$m_s/m_b$ to be smaller than $m_\mu/m_\tau$.  From 
(\ref{Umasses})-(\ref{Lmasses}) one sees that both these assertions are 
correct.  Further, quarks being heavier than leptons and therefore closer 
to the fixed point $(1,0,0)$ where rotation is slow, will naturally also 
have smaller mixing than leptons, which is again seen in (\ref{CKMmat}) 
and (\ref{MNSmat}) to be clearly borne out by experiment .  One can even 
go to more details, and explain the relative sizes of elements within each 
mixing matrix \cite{features}.  To a good approximation, the state vectors of 
the three generations can be represented as a orthonormal (Darboux) triad at 
the location of the heaviest generation as illustrated in Figure \ref{Darboux},
with the heaviest generation state as the radial vector to the sphere,
the second generation state as the tangent vector to the trajectory, and
the lightest generation state as the vector orthogonal to both the above.
The mixing matrix then appears just as the matrix representing the rotation 
undergone by this triad as it is transported along the trajectory from 
the location of the heaviest up-state to the heaviest down-state.  To
leading order in the distance transported, elementary differential geometry
\cite{Docarmo} gives this rotation matrix as:
\begin{equation}
V_{CKM} \sim \left( \begin{array}{ccc}
   1 & - \kappa_g \Delta s & - \tau_g \Delta s \\
   \kappa_g \Delta s & 1 & \kappa_n \Delta s \\
   \tau_g \Delta s & - \kappa_n \Delta s & 1  \end{array} \right),
\label{Muimat}
\end{equation}
with $\kappa_n$ being the normal curvature, $\kappa_g$ the geodesic curvature,
and $\tau_g$ the geodesic torsion of a curve on a surface.  For our unit
sphere, $\kappa_n = 1$ and $\tau_g = 0$.  From this we deduce first that the
corner elements (13 and 31) are of second order in $\Delta s$ and therefore
small compared with the others, which they are in experiment for both 
quarks and leptons as seen in (\ref{CKMmat}) and (\ref{MNSmat}).  Secondly,
we conclude that the 23 and 32 elements are given approximately just by
the transportation distance $\Delta s$, namely for the quark case by the
distance between the top and bottom quarks along the trajectory, and for
the lepton case by the distance between $\tau$ and $\nu_3$.  And indeed, if 
one takes the trouble to measure with a bit of string these distances on
the the trajectory in Figure \ref{Jakovsphere}, one will find values very
close to the experimental numbers given for these elements in (\ref{CKMmat})
and (\ref{MNSmat}).

Of course, having actually done the calculation, one can make much a more
detailed comparison of the result with experiment than that afforded by the
above qualitative estimates.  Indeed, from the calculation \cite{phenodsm}, 
one obtains the numbers given in Table \ref{CKMtable},
\begin{table}
\begin{eqnarray*}
\begin{array}{||c||c||c|c||}  
\hline \hline
Quantity & Experimental Range & Predicted & Predicted Range \\
         &                    & Central Value &             \\
\hline \hline
|V_{ud}| & 0.9745 - 0.9760 & 0.9753 & 0.9745 - 0.9762 \\ \hline
|V_{us}| & 0.217 - 0.224 & (0.2207) & input \\ \hline
|V_{ub}| & 0.0018 - 0.0045 & 0.0045 & 0.0043 - 0.0046 \\ \hline
|V_{cd}| & 0.217 - 0.224 & (0.2204) & input \\ \hline
|V_{cs}| & 0.9737 - 0.9753 & 0.9745 & 0.9733 - 0.9756 \\ \hline
|V_{cb}| & 0.036 - 0.042 & 0.0426 & 0.0354 - 0.0508 \\ \hline
|V_{td}| & 0.004 - 0.013 & 0.0138 & 0.0120 - 0.0157 \\ \hline
|V_{ts}| & 0.035 - 0.042 & 0.0406 & 0.0336 - 0.0486 \\ \hline
|V_{tb}| & 0.9991 - 0.9994 & 0.9991 & 0.9988 - 0.9994 \\ \hline
|V_{ub}/V_{cb}| & 0.08 \pm 0.02 & 0.1049 & 0.0859 - 0.1266 \\ \hline
|V_{td}/V_{ts}| & < 0.27 & 0.3391 & 0.3149 - 0.3668 \\ \hline
|V_{tb}^{*}V_{td}| & 0.0084 \pm 0.0018 & 0.0138 & 0.0120 - 0.0156 \\ \hline
   \hline
|U_{\mu3}| & 0.56 - 0.83 & 0.6658 & 0.6528 - 0.6770 \\ \hline
|U_{e3}| & 0.00 - 0.15 & 0.0678 & 0.0632 - 0.0730 \\ \hline
|U_{e2}| & 0.4 - 0.7 & 0.2266 & 0.2042 - 0.2531 \\ \hline \hline 
\end{array}
\end{eqnarray*}
\caption{Predicted CKM matrix elements for both quarks and leptons}
\label{CKMtable}
\end{table} 
where one sees that all entries more or less overlap with the present
experimental limits, except for the solar neutrino mixing element $U_{e2}$,
which being related to the trajectory-dependent geodesic curvature according 
to (\ref{Muimat}) is particular difficult for our calculation to get correct.
We note that all these numbers have been obtained by adjusting only one 
parameter to the Cabibbo angle $V_{us} \sim V_{cd}$, the other two parameters 
in the calculation having already been fitted to fermion masses.  Thus,
unless this agreement with experiment turns out to be all fortuitous,
it would appear that starting with the identification of dual colour to
generation, one can indeed explain not only that there are three and only
three generations of fermions but that they have have the experimentally
observed mass and mixing patterns, namely all the features set out at the
beginning.  

However, the problem does not stop there.  Given that new physical assumptions
have been made, new consequences will follow so that one will need first to 
ensure that these do not contradict present experiment, and second, to see 
whether they lead to predictions testable in the not too distant future.  As 
always, this is a lengthy business, which for the suggested scheme is still 
far from complete.  

Indeed, the only area which has so far been explored in some detail is 
the exchange of dual colour gauge bosons which is bound to occur when 
$\widetilde{SU}(3)$ is a local gauge symmetry.  Firstly, these bosons 
carrying dual colour, i.e. generation index, but no electrical charge,
can lead, when exchanged, to flavour-changing neutral current (FCNC) 
effects.  Secondly, they can be exchanged between any fermions carrying 
a generation index including in particular neutrinos, an thus give rise 
to new interactions hitherto unsuspected.  The size of both these two 
types of effects depends on the mass of the dual colour bosons which 
unfortunately is not given by the theory, and is also left undetermined by 
the above fit to the fermion mass and mixing parameters.  However, the 
coupling strength of the dual colour gauge bosons is given in terms of 
that of the colour gauge bosons by the Dirac quantization condition, while
the branching of this coupling to various modes by the above calculation
of the fermion mixing matrices.  Hence, once given a value for the scale
of the dual colour gauge boson mass, the above scheme will give detailed 
predictions for all processes due to the exchange of one of these bosons
\cite{fcnc}.  The absence of any observed signal at present of any FCNC effect
puts a lower bound on the dual colour gauge boson mass.  The strongest bound
was obtained from the $K_L - K_S$ mass difference giving a scale for the dual 
colour gauge boson mass of order 400 TeV \cite{fcnc}.  What is more relevant,
however, would be an upper bound on this mass, which would predict the
level at which FCNC effects will occur.  This is usually not available
in models of horizontal symmetries and it is thus quite interesting 
that a possibility for deriving such a bound in the present scheme is 
seen to arise from a quite unexpected direction \cite{airshowers}, as follows. 

As already explained, even neutrinos are expected to acquire a new, and 
by Dirac quantization condition, strong interaction from dual colour gauge 
boson exchange.  This prediction is at first sight frightening until one
recalls that, from the above estimate for the dual colour gauge boson mass,
this interaction will not manifest itself until c.m. energies above 400 
TeV.  This is way above any energy achievable in the laboratory in the
foreseeable future, or in any known astrophysical phenomenon apart from
one notable exception, namely EHECR (extremely high energy cosmic rays).
For a cosmic ray primary hitting a nucleon in an air nucleus, 400 TeV in
the c.m. means an incoming energy of roughly $5\times 10^{19}$eV, and some 
dozens of rare air shower events above this energy have been observed 
over the last few decades \cite{Takeda}.  To cosmic ray physicists, these 
events are a headache in that protons and nuclei of such energy beyond 
the so-called Greisen-Zatsepin-Ku'zmin (GZK) \cite{Greisemin} cut-off are 
not supposed to survive a long journey through the 2.7 K microwave background, 
and there are no nearby celestial bodies likely to produce particles of this
sort of energies.  A possible solution is that these post-GZK air showers are 
due not to protons but to some neutral particle which does not interact with 
the microwave background, but the only stable neutral particle we know is the 
neutrino (the photon being ruled out by other considerations), and neutrinos 
with only weak interactions are incapable of producing air showers with the 
frequency and distributions seen.  However, if neutrinos can acquire at these 
high energies a strong interaction as predicted by the present scheme, then
they can both survive a long journey through the microwave background and
produce post-GZK air showers as observed \cite{airshowers}.  Accepting this 
as possible solution to the GZK puzzle yields then a rough upper bound to 
the dual colour gauge boson mass \cite{fcnc}.

The conclusion of the analysis to-date of dual colour gauge boson exchange 
is thus as follows.  So far no violation of experimental bounds have been
found in either low energy FCNC processes or high energy neutrino reactions.
Instead, one gains a possible explanation for the old cosmic ray puzzle of
post-GZK air showers and an upper bound on the dual colour gauge boson
mass, which yields in turn quantitative predictions for FCNC effects.  Of 
such predictions thus obtained, the most interesting are the rate of the 
rare decay $K_L \rightarrow \mu^{\pm} e^{\mp}$, the mass difference between 
the neutral charmed mesons $D_0 - \bar{D}_0$ \cite{fcnc} and the rate of
the coherent conversion of $\mu$ to $e$ in nuclei such as $Al$ and $Sn$ 
\cite{mueconv}, all of which are already quite close to the present 
experimental bounds.  Tests on these predictions, however, cannot be made 
very decisive at present since they depend on the dual colour gauge boson 
mass to the 4th power, which mass is but poorly estimated by the scanty 
data on post-GZK air showers, even if one accepts our explanation there.  
With more data in the near future \cite{Auger} and a more careful analysis, 
however, the situation can be improved.

There are other areas where the present scheme makes some quite novel
predictions testable by experiment, which we have been studying but are 
not yet in the position to decribe in detail.\footnote{In fact, since 
the meeting, two further papers \cite{impromat,photrans} have appeared 
examining the scheme's predictions on a new class of phenomena called
fermion ``transmutation'' which occurs as a consequence of the mass 
matrix rotation.}

In summary, I would say that, up to the present, Yang-Mills duality does 
seem to offer a viable solution to the generation puzzle, besides predicting
some very new physicial phenomena which will be interesting to explore in 
the future.

\end{document}

%% file: loopcurv.pstex_t
\begin{picture}(0,0)%
\epsfig{file=loopcurv.pstex}%
\end{picture}%
\setlength{\unitlength}{3947sp}%
\begingroup\makeatletter\ifx\SetFigFont\undefined%
\gdef\SetFigFont#1#2#3#4#5{%
  \reset@font\fontsize{#1}{#2pt}%
  \fontfamily{#3}\fontseries{#4}\fontshape{#5}%
  \selectfont}%
\fi\endgroup%
\begin{picture}(6774,1980)(1609,-3391)
\end{picture}